\documentclass[aps,prd,superscriptaddress,preprint,tightenlines,nofootinbib]{revtex4}

\usepackage{graphicx} 
\usepackage{dcolumn}  
\usepackage{bm}       
\usepackage{epstopdf} 
\begin{document}

\preprint{CLNS 08/2042}       
\preprint{CLEO 08-24}         

\title{\boldmath First model-independent determination of the relative strong phase between $D^0$ and $\bar D^0$ $\to K^0_S\pi^+\pi^-$
and its impact on the CKM Angle $\gamma/\phi_3$ measurement}



\author{R.~A.~Briere}
\author{H.~Vogel}
\affiliation{Carnegie Mellon University, Pittsburgh, Pennsylvania 15213, USA}
\author{P.~U.~E.~Onyisi}
\author{J.~L.~Rosner}
\affiliation{Enrico Fermi Institute, University of
Chicago, Chicago, Illinois 60637, USA}
\author{J.~P.~Alexander}
\author{D.~G.~Cassel}
\author{J.~E.~Duboscq}\thanks{Deceased}
\author{R.~Ehrlich}
\author{L.~Fields}
\author{L.~Gibbons}
\author{R.~Gray}
\author{S.~W.~Gray}
\author{D.~L.~Hartill}
\author{B.~K.~Heltsley}
\author{D.~Hertz}
\author{J.~M.~Hunt}
\author{J.~Kandaswamy}
\author{D.~L.~Kreinick}
\author{V.~E.~Kuznetsov}
\author{J.~Ledoux}
\author{H.~Mahlke-Kr\"uger}
\author{D.~Mohapatra}
\author{J.~R.~Patterson}
\author{D.~Peterson}
\author{D.~Riley}
\author{A.~Ryd}
\author{A.~J.~Sadoff}
\author{X.~Shi}
\author{S.~Stroiney}
\author{W.~M.~Sun}
\author{T.~Wilksen}
\affiliation{Cornell University, Ithaca, New York 14853, USA}
% ...
\collaboration{CLEO Collaboration}
\noaffiliation


\date{March 9, 2009}

\begin{abstract} 
We exploit the quantum coherence between pair-produced $D^0$ and $\bar D^0$ in 
$\psi(3770)$ decays to make a first determination of the relative strong phase differences
between $D^0\to K_S^0\pi^+\pi^-$ and $\bar D^0\to K_S^0\pi^+\pi^-$, 
which are of great importance in determining
the CKM angle $\gamma/\phi_3$ in $B^-\to D^0 (\bar D^0) K^-$ decays.
Using 818 pb$^{-1}$ of $e^+ e^-$ collision data 
collected with the CLEO-c detector at $E_{\rm cm}= 3.77$ GeV, we employ
a binned Dalitz-plot analysis 
of $K^0_S\pi^+\pi^-$ and $K^0_L\pi^+\pi^-$ decays recoiling against
flavor-tagged, CP-tagged and $K^0_S\pi^+\pi^-$ tagged events 
to determine these strong phase differences.
\end{abstract}

\maketitle

\section{Introduction}
A central goal of flavor physics is the determination of all elements of the CKM matrix~\cite{CKM},
magnitudes and phases. Of the three angles of the $b-d$ CKM triangle, denoted $\alpha$,
$\beta$, and $\gamma$ by some, $\phi_2$, $\phi_1$, and $\phi_3$ by others, the least-well
determined is $\gamma/\phi_3$, the phase of $V_{ub}$ relative to $V_{cb}$. It is of great interest
to determine $\gamma/\phi_3$ using the decay $B^{\pm}\to K^{\pm}\tilde D^0$, since in this
mode, the $\gamma/\phi_3$ value obtained
is expected to be insensitive to new physics effects in $B$ decays.
Here, $\tilde D^0$ is either $D^0$ or $\bar D^0$, and both decay to the same final state, and 
so their amplitudes add.
Sensitivity to the angle $\gamma/\phi_3$ comes from the interference between two
Cabibbo-suppressed diagrams: $b\to c\bar u s$, giving 
rise to $B^-\to K^-D^0$, and the color and CKM suppressed process $b\to u\bar c s$, giving rise to $B^-\to K^-\bar D^0$.
One of the most promising $\tilde D^0$ decays for measuring $\gamma/\phi_3$ using this method
is $\tilde D^0 \to K^0_S\pi^+\pi^-$,
because it is Cabibbo favored (CF) for both
$D^0$ and $\bar{D}^0$ decays, thus providing large event yields. To make use of this decay, however,
the interference effects between $B^-\to K^- \bar D^0(\to K^0_S\pi^+\pi^-)$ and
$B^-\to K^- D^0(\to K^0_S\pi^+\pi)$ need to be understood. These interference effects can
be understood and measured using CLEO-c data.

We first write the amplitude for the $B^{\pm}$ decay as follows:

\begin{equation}
\mathcal A(B^{\pm}\to K^{\pm}\tilde D^0, \tilde D^0\to K_S^0\pi^+\pi^-(x,y)) 
\propto f_D(x,y) + r_Be^{i\theta_{\pm}} f_{\bar D}(x,y).
\label{eq:amp1}
\end{equation}

\noindent Here, $x\equiv m^2_{K_S^0\pi^+}$, $y\equiv m^2_{K_S^0\pi^-}$ are the Dalitz-plot variables in the 
$\tilde D^0$ decay, $f_D(x,y) (f_{\bar D}(x,y))$ is the amplitude for $D^0 (\bar D^0)$ decay 
to $K_S^0\pi^+\pi^-$ at $(x,y)$, 
$r_B$ is the ratio of the suppressed to favored amplitudes, and 
$\theta_{\pm}\equiv \delta_B\pm \gamma$, where $\delta_B$ is the strong phase shift between the
color-favored and color-suppressed amplitudes. Ignoring the second-order effects
of charm mixing and CP violation~\cite{GGSZ,GSZ}, we have $f_{\bar{D}}(x,y) = f_D(y,x)$, and
Eq.~\ref{eq:amp1} can then be rewritten as:

\begin{equation}
\mathcal A(B^{\pm}\to K^{\pm}\tilde D^0, \tilde D^0\to K_S^0\pi^+\pi^-(x,y))
\propto f_D(x,y) + r_Be^{i\theta_{\pm}}f_D(y,x).
\end{equation}

\noindent The square of the amplitude clearly depends on the phase difference
$\Delta \delta_D\equiv \delta_D(x,y)-\delta_D(y,x)$,
where $\delta_D(x,y)$ is the phase of $f_D(x,y)$. Thus, for the determination of $\gamma/\phi_3$,
one must know $\Delta \delta_D(x,y)$.

Previous analyses extracted $\Delta \delta_D(x,y)$ by fitting a flavor-tagged 
$D^0\to K_S^0\pi^+\pi^-$ Dalitz plot to a model for $D^0$ decay involving various
2-body intermediate states \cite{GGSZ_BaBar1,GGSZ_BaBar2,GGSZ_Belle}. Such an approach
introduces a $7^{\circ}\sim 9^{\circ}$ model uncertainty in the value of $\gamma/\phi_3$,
\footnote{BaBar claims $5^{\circ}$ uncertainty on $\gamma/\phi_3$ in \cite{GGSZ_BaBar2}
by combining $K_S^0K^+K^-$ and $K_S^0\pi^+\pi^-$ modes.}
which would be a limiting uncertainty for LHCb \cite{Libby} and future $B$-factory experiments. 

In the analysis presented here, we employ a model-independent approach to obtain 
$\Delta \delta_D(x,y)$ as suggested by Giri {\it et al.} \cite{GGSZ}, by exploiting 
the quantum coherence of $D^0-\bar D^0$ pairs at the $\psi(3770)$. Because of this quantum
correlation, $K_S^0\pi^+\pi^-$ and $K_L^0\pi^+\pi^-$ decays recoiling against flavor tags,
CP-tags, and $D^0\to K_S^0\pi^+\pi^-$ tags, taken together provide direct sensitivity to 
the quantities $\cos\Delta\delta_D$ and $\sin\Delta\delta_D$. This measurement will result
in a substantial reduction in the systematic uncertainty associated with the 
interference effects between $B^-\to K^- \bar D^0(\to K^0_S\pi^+\pi^-)$ and
$B^-\to K^- D^0(\to K^0_S\pi^+\pi)$.
 
\section{Formalism}
\label{section:diff_KsKl}
Giri {\it et al.} proposed \cite{GGSZ} a model-independent procedure for obtaining 
$\Delta\delta_D(x,y)$, as follows. The Dalitz plot is divided into $2\mathcal N$ bins,
symmetrically about the line $x=y$. The bins are indexed from $-i$ to $i$, excluding 
zero. The coordinate exchange 
$x\leftrightarrow y$ thus corresponds to the exchange of bins $i\leftrightarrow -i$.
The number of events in the $i$-th bin of a flavor-tagged $K_S^0\pi^+\pi^-$ Dalitz plot 
from a $D^0$ decay is then expressed as:
 
\begin{equation} 
K_i = A_D\int_{i}|f_D(x,y)|^2dxdy = A_DF_i,
\label{eq:ki}
\end{equation}

\noindent 
where $A_D$ is a normalization factor. 
The interference between the $D^0$ and $\bar D^0$ amplitudes is parameterized by two 
quantities

\begin{equation} 
c_i \equiv \frac{1}{\sqrt{F_iF_{-i}}} \int_i|f_D(x,y)||f_D(y,x)|\cos[\Delta\delta_D(x,y)]dxdy,
\label{eq:ci}
\end{equation}

\noindent and 

\begin{equation} 
s_i \equiv \frac{1}{\sqrt{F_iF_{-i}}} \int_i |f_D(x,y)||f_D(y,x)|\sin[\Delta\delta_D(x,y)]dxdy,
\end{equation}

\noindent where the integral is performed over a single bin.
The parameters $c_i$ and $s_i$ are the amplitude-weighted averages of $\cos{\Delta\delta_{D}}$ and $\sin{\Delta\delta_{D}}$ over each Dalitz-plot bin.
It is important to note that $c_i$
and $s_i$ depend only on the $D^0$ decay, not the $B$ decay, and therefore these quantities can be 
measured using CLEO-c data. In principle they could be left as free parameters in a 
$\tilde D^0\to K_S^0\pi^+\pi^-$ Dalitz-plot analysis from $B^{\pm}$ decays, but their values
can be more precisely determined from correlated $D^0\bar D^0$ pairs produced in CLEO-c.             

Though the original idea of Giri {\it et al.} was to divide the Dalitz plot into square
bins \cite{GGSZ}, Bondar {\it et al.} noted \cite{Bondar} that increased
sensitivity is obtained if the bins are chosen to minimize the variation in $\Delta\delta_D$
over each bin. Thus, we 
divide the Dalitz phase space into ${\mathcal N}$ bins of equal size with respect to $\Delta\delta_{D}$ as predicted by the 
BaBar isobar model \cite{GGSZ_BaBar1}. 
In the half of the Dalitz plot $m^2(K_S^0\pi^+)<m^2(K_S^0\pi^-)$, the $i^{th}$ bin is
defined by the condition 

\begin{equation} 
2\pi(i-3/2)/{\mathcal N} < \Delta \delta_D(x,y) < 2\pi(i-1/2)/{\mathcal N},
\end{equation} 

\begin{figure}
\centering
\includegraphics[width=8cm]{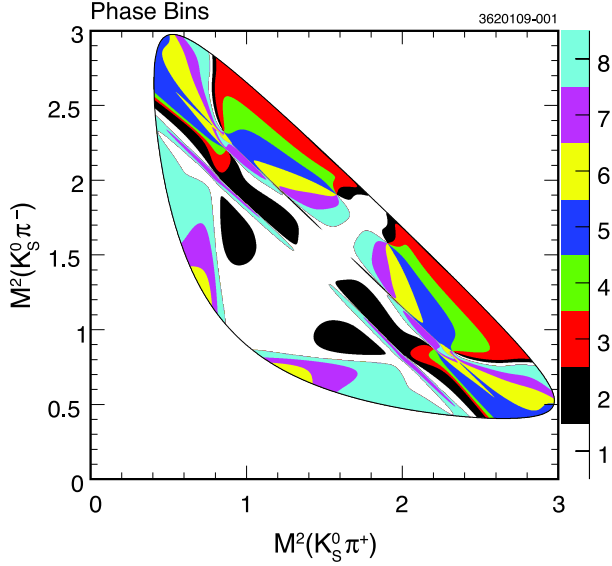}
\caption{Phase binning of the $D^0 \to \bar K_S^0 \pi^+\pi^-$ Dalitz plot.}
\label{fig:phasebin}
\end{figure}

\noindent The $-i^{th}$ bin is defined symmetrically in the lower portion of the Dalitz plot. 
Such a binning with $\mathcal N=8$ is shown in Fig.~\ref{fig:phasebin}.
One might suspect that because we are using a model to determine our bins, we are not 
free of model dependence. In fact {\it any} binning is correct in that it will
give a correct, unbiased answer for $\gamma/\phi_3$, at the cost of larger 
uncertainties compared to an optimal binning with respect to $\Delta\delta_{D}$.

We now describe how CLEO-c data can be used to determine $c_i$ and $s_i$.
The event yields in the $i^{th}$ bin of both flavor-tagged and CP-tagged
$\tilde D^0\to K^0_S\pi^+\pi^-$ Dalitz plot are required.
Because the $\psi(3770)$ has $C=-1$, the CP of the $\tilde D^0\to K^0_S\pi^+\pi^-$ decay can be 
determined by reconstructing the companion $\tilde D^0$ in a CP eigenstate. With a CP-tagged
$\tilde D^0\to K^0_S\pi^+\pi^-$ decay, the amplitude is given by:

\begin{equation} 
f_{CP\pm}(x,y) = \frac{1}{\sqrt{2}}[f_D(x,y)\pm f_D(y,x)],
\label{eqn:Amp_CP_Ks}
\end{equation} 

\noindent for CP-even and CP-odd states of a $\tilde D^0\to K^0_S\pi^+\pi^-$ decay. 
Since the event rate is proportional 
to the square of this amplitude, the number of events in the $i^{th}$ bin of a CP-tagged Dalitz plot is then:

\begin{equation} 
M_i^{\pm}=h_{CP\pm}(K_i\pm 2c_i\sqrt{K_iK_{-i}}+K_{-i}),
\label{eqn:CP_Ks}
\end{equation}

\noindent 
\noindent where $h_{CP\pm} = S^{\pm}/2S_f$ is a normalization factor that
depends on the number, $S_f$, of 
single flavor-tagged signal decays, and the number, $S^{\pm}$, of single CP-tagged signal decays.
Thus, access to $c_i$ is enabled by measuring
the number of events, $M_i^{\pm}$, in a CP-tagged $K_S^0\pi^+\pi^-$ Dalitz plot,
and the number of events, $K_i$, in a flavor-tagged $K_S^0\pi^+\pi^-$ Dalitz plot.

Unfortunately, as evident from Eq.~\ref{eq:ci}, the sign of $\Delta\delta_D$ is undetermined in
each of the $i$ bins. However, sensitivity to both $c_i$ and $s_i$ can be obtained by analyzing
$D^0\to K_S^0\pi^+\pi^-$ vs. $\bar D^0 \to K_S^0\pi^+\pi^-$ data. The amplitude
for $\psi(3770)$ decays to two $K_S^0\pi^+\pi^-$ decays is as follows:

\begin{equation} 
f(x,y,x',y')=\frac{f_D(x,y)f_D(y',x')-f_D(x',y')f_D(y,x)}{\sqrt{2}}.
\label{eqn:Amp_Ks_Ks}
\end{equation}

\noindent The primed and unprimed Dalitz-plot coordinates correspond to the Dalitz-plot variables 
of the two $\tilde D^0\to K^0_S\pi^+\pi^-$ decays.
Defining $M_{ij}$ as the event rate in the $i^{th}$ bin of the first and the $j^{th}$ bin of the
second $\tilde D^0\to K^0_S\pi^+\pi^-$ Dalitz plots, respectively, we have:

\begin{eqnarray} 
M_{ij}&=&h_{corr}(K_iK_{-j}+K_{-i}K_j- 
 2\sqrt{K_iK_{-j}K_{-i}K_j}(c_ic_j+s_is_j)).
\label{eqn:Ks_Ks}
\end{eqnarray}

\noindent Here, $h_{corr} = N_{D\bar{D}}/2S_f^2$, where 
$N_{D\bar{D}}$ is
the number of $D\bar{D}$ pairs, and as before $S_f$ is the number of flavor-tagged signal decays.
Equation~\ref{eqn:Ks_Ks} then relates the product
$(c_i c_j + s_i s_j)$ to the measured yields of events in the 
flavor-tagged $\tilde D^0\to K_S^0\pi^+\pi^-$ Dalitz plot ($K_{i,j}$'s) and the yields
in the $D^0\to K_S^0\pi^+\pi^-$ vs. $\bar D^0 \to K_S^0\pi^+\pi^-$ ($M_{ij}$'s) Dalitz plots.
The sensitivity to this product leads to a four-fold ambiguity: change of sign of 
all $c_i$ or all $s_i$. In combination with
the CP-tagged analysis though, where the sign of $c_i$ is determined, this reduces to a
two-fold ambiguity. One of the two solutions can be chosen based on a weak model assumption \cite{GGSZ_BaBar1}.

The decay $D^0\to K_L^0\pi^+\pi^-$, due to its close relationship with $D^0\to K_S^0\pi^+\pi^-$,
can be used to further improve the $c_i$ and $s_i$ 
determination. Since the $K_S^0$ and $K_L^0$ mesons are of opposite CP, and we 
assume the convention that $A(D^0\to K_S^0\pi^+\pi^-)=A(\bar D^0\to K_S^0\pi^-\pi^+)$,
it then follows that $A(D^0\to K_L^0\pi^+\pi^-)=-A(\bar D^0\to K_L^0\pi^-\pi^+)$. 
Then, for $K^0_L\pi^+\pi^-$, the Dalitz-plot rates of Eq.~\ref{eqn:CP_Ks} (CP vs. $D^0\to K_S^0\pi^+\pi^-$) and Eq.~\ref{eqn:Ks_Ks} ($D^0\to K_S^0\pi^+\pi^-$ vs. $\bar D^0\to K_S^0\pi^+\pi^-$) become:

\begin{eqnarray} 
\label{eqn:CP_Kl}
M_i^{\pm}&=&h_{CP\pm}(K_i'\mp 2c_i'\sqrt{K_i'K_{-i}'}+K_{-i}'),\\
M_{ij}&=&h_{corr}[K_iK_{-j}'+K_{-i}K_j'+
2\sqrt{K_iK_{-j}'K_{-i}K_j'}(c_ic_j'+s_is_j')],
\label{eqn:Ks_Kl}
\end{eqnarray}

\noindent for CP vs. $D^0\to K_L^0\pi^+\pi^-$  and $D^0\to K_S^0\pi^+\pi^-$ vs. $\bar D^0\to K_L^0\pi^+\pi^-$, respectively, where $c_i'$, $s_i'$ are associated with $D^0\to K_L^0\pi^+\pi^-$ decay.

For $D^0\to K_L^0\pi^+\pi^-$ decays to benefit our determination of $c_i$ and $s_i$, 
we must determine the differences $\Delta c_i\equiv c_i'-c_i$, and $\Delta s_i \equiv s_i'-s_i$.
In addition to the relative sign change in Eq.~\ref{eqn:CP_Kl} and 
Eq.~\ref{eqn:Ks_Kl}, doubly Cabibbo suppressed decays (DCSD) of $D^0/\bar D^0$ also
contribute with opposite signs in $D^0\to K_S^0\pi^+\pi^-$ and $D^0\to K_L^0\pi^+\pi^-$
decays. We can see this by inspecting the $D^0$ decay 
amplitude for each Dalitz plot

\begin{eqnarray} 
A(K_S^0\pi^+\pi^-) = \frac{1}{\sqrt{2}}[A(K^0\pi^+\pi^-)+A(\bar K^0\pi^+\pi^-)],\\
A(K_L^0\pi^+\pi^-) = \frac{1}{\sqrt{2}}[A(K^0\pi^+\pi^-)-A(\bar K^0\pi^+\pi^-)].
\end{eqnarray}

\noindent The effect of this relative minus sign is to introduce a $180^\circ$ phase difference for 
all DCSD $K^*$ resonances in the $K_L^0\pi^+\pi^-$ model. We can use U-spin symmetry
to relate the amplitudes for resonances of definite CP eigenvalue, e.g. $K_{S,L}^0\rho^0(770)$. 
We find that 
these states acquire a factor of $re^{i\delta}$. To convert a $D^0\to K_S^0\pi^+\pi^-$ model to
the corresponding $D^0\to K_L^0\pi^+\pi^-$ model, we multiply all DCSD 
amplitudes by $-1$ and multiply each CP eigenstate amplitude by $(1-2re^{i\delta})$, with
$r=\tan^2\theta_C$ and $\delta=0^\circ$ fixed for every resonance (here $\theta_C$ is the 
Cabibbo angle).  
We then determine central values for the corrections, $\Delta c_i\equiv c_i'-c_i$, $\Delta s_i\equiv s_i'-s_i$
using the $D^0 \to K_S^0\pi^+\pi^-$ BaBar model~\cite{GGSZ_BaBar1}.
We ascribe uncertainty to both the choice of $r$ and $\delta$, as well as the usage
of the BaBar model to determine the uncertainties on $\Delta c_i$ and $\Delta s_i$. The former
are estimated by varying the phase $\delta$ between 0 and $2\pi$, and $r$ 
by $\pm50\%$. For the latter, we estimate
$\Delta c_i$ and $\Delta s_i$ using the Belle~\cite{GGSZ_Belle} and CLEO~\cite{David} 
$D^0\to K_S^0\pi^+\pi^-$ isobar model fits, and take the largest resulting deviation 
from the value found with the BaBar model as a model-dependent systematic uncertainty. 
The total systematic uncertainties in the 
corrections are the quadrature sum of these two uncertainties. The central values
and uncertainties on $\Delta c_i$ and $\Delta s_i$ 
are shown in Table~\ref{table:correction}.

\begin{table}[t]
\centering
\caption{Predicted values for $\Delta c_i$ and $\Delta s_i$ with the systematic uncertainties.}
\begin{tabular}{c|cc}        \hline\hline
$i$ & $\Delta c_i$ & $\Delta s_i$ \\ \hline
0 & $0.099  \pm  0.040$ & $-0.034  \pm  0.068$ \\
1 & $0.167  \pm  0.029$ & $-0.064  \pm  0.084$ \\
2 & $0.327  \pm  0.122$ & $-0.013  \pm  0.097$ \\
3 & $0.253  \pm  0.192$ & $ \hphantom{-}0.133  \pm  0.136$ \\
4 & $0.077  \pm  0.061$ & $ \hphantom{-}0.041  \pm  0.080$ \\
5 & $0.220  \pm  0.084$ & $-0.038  \pm  0.065$ \\
6 & $0.416  \pm  0.160$ & $ \hphantom{-}0.095  \pm  0.063$ \\
7 & $0.184  \pm  0.024$ & $ \hphantom{-}0.015  \pm  0.086$ \\
\hline\hline
\end{tabular}
\label{table:correction}
\end{table}

\section{Event Selection}
We analyze 818 pb$^{-1}$ of $e^+ e^-$ collision data produced by the Cornell Electron
Storage Ring (CESR) at $E_{\rm cm}=3.77$ GeV and collected with the CLEO-c detector. 
The CLEO-c detector is a general purpose 
solenoidal detector which includes a tracking system for measuring momentum and specific
ionization ($dE/dx$) of charged particles, a Ring Imaging Cherenkov detector (RICH) to
aid in particle identification, and a CsI calorimeter for detection of electromagnetic
showers. The CLEO-c detector is described in detail elsewhere~\cite{CLEO}.

    Standard CLEO-c selection criteria for $\pi^{\pm}$, $K^{\pm}$, $\pi^0$, and $K_S^0$ 
candidates are used, and are described in Ref.~\cite{DHadronic}. To distinguish
electrons from hadrons, we use a multivariate discriminant~\cite{EID}
that combines information from the ratio of the energy deposited in the calorimeter to 
the measured track momentum ($E/p$), ionization energy loss in the tracking chamber ($dE/dx$),
and the ring-imaging Cherenkov counter (RICH). For $K_S^0$ decays,
we select candidates with $|M(\pi^+\pi^-)-M_{K_S^0}|<7.5$ MeV$/c^2$, and require the decay vertex to be
separated from the interaction region with a significance greater than two standard 
deviations (except for $D^0\to K_S^0\pi^+\pi^-$ vs. $\bar D^0\to K_S^0\pi^+\pi^-$ candidates). 
Reconstruction of $\eta\to\gamma\gamma$ proceeds analogously to $\pi^0\to\gamma\gamma$,
with the requirement that $|M(\gamma\gamma)-M_{\eta}|<42$ MeV$/c^2$.
We form $\omega\to\pi^+\pi^-\pi^0$ candidates and require their mass to be within 20 MeV
of the nomimal $\omega$ mass~\cite{PDG}.

	In this analysis, we reconstruct $D^0$ mesons in several flavor-tagged modes,
CP-tagged modes, and in $K^0_S\pi^+\pi^-$. From these selected events, we also reconstruct
the companion $D^0$ from the $\psi(3770)$ decay in either $K^0_S\pi^+\pi^-$ or $K^0_L\pi^+\pi^-$ 
to form ``double-tags''. The
single tags yields enter our analysis through the $S_f$ and $S^{\pm}$ factors,
whereas the double-tags provide the $K_i$, $M_i$ and $M_{ij}$ yields across their
respective Dalitz plots. The double-tagged events we consider are shown in Table~\ref{table:modes} 
(all the notations include charge conjugate if not otherwise specified.).
We thus consider flavor tags: $K^-\pi^+$, $K^-\pi^+\pi^0$, $K^-\pi^+\pi^+\pi^-$;
semileptonic tag: $K^-e^+\nu$; CP-even tags: $K^+K^-$, $\pi^+\pi^-$, $K_S^0\pi^0\pi^0$, $K_L^0\pi^0$;
and CP-odd tags: $K_S^0\pi^0$, $K_S^0\eta$, $K_S^0\omega$, 
with $\pi^0/\eta\to \gamma\gamma$, $\omega\to \pi^+\pi^-\pi^0$,
and $K_S^0\to \pi^+\pi^-$. We also reconstruct double-tag (DT) events with 
$\tilde D^0\to K^0_{S,L}\pi^+\pi^-$ vs $\tilde D^0\to K^0_S\pi^+\pi^-$ as discussed in
the preceding section. We do not reconstruct $K_L^0\pi^+\pi^-$ in some DT modes when
there are two missing particles ($K^-e^+\nu$, and $K_L^0\pi^0$ cases) or 
the backgrounds are large (as for $K_S^0\pi^0\pi^0$, and $K_S^0\omega$).

\begin{figure}
\centering
\includegraphics[width=8cm]{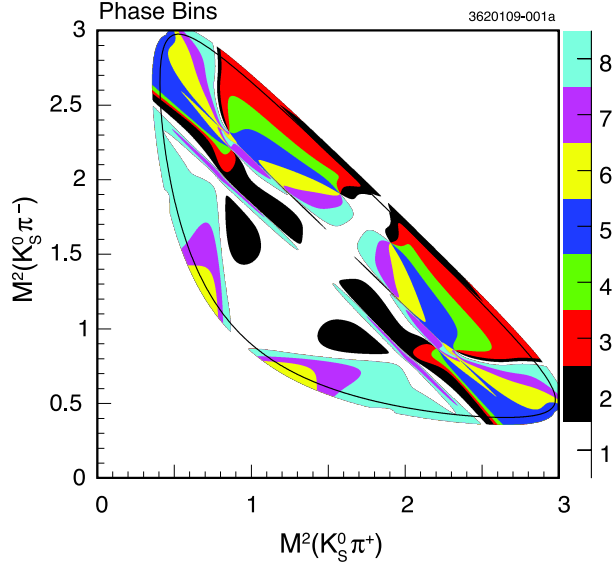}
\caption{Binning of the $D^0 \to \bar K_S^0 \pi^+\pi^-$ Dalitz plot with respect to $\Delta\delta_{D}$. 
The bins are extended beyond the kinematically allowed Dalitz-plot boundary.}
\label{fig:phasebin-smear}
\end{figure}

The event yield in the $i^{th}$ Dalitz-plot bin of each tagged $\tilde D^0\to K^0_{S,L}\pi^+\pi^-$ sample
is determined by evaluating the phase difference for each data point according to the BaBar isobar model. 
The contribution of each isobar to the total amplitude is evaluated
as a function of all three invariant mass-squared combinations computed directly from the four-momentum of the $\tilde D$ daughters as described in Ref.~\cite{Kopp:2000gv}. 
The phase difference is well defined beyond and continues smoothly across the kinematically allowed
Dalitz-plot boundary as shown in Fig.~\ref{fig:phasebin-smear}. A small number of candidate events ($\sim$1-3\% depending on tag and signal mode) included in this analysis are reconstructed outside the kinematically allowed region due to finite detector resolution.

\begin{table}[b]
\centering
\caption{Reconstructed Double Tag modes.}
\begin{tabular}{lcc}        \hline\hline
 Mode  & $K_S^0\pi^+\pi^-$ & $K_L^0\pi^+\pi^-$\\\hline
\multicolumn{3}{c}{Flavor Tags} \\
\hline
$K^-\pi^+$  & $\times$ & $\times$\\
$K^-\pi^+\pi^0$ & $\times$ & $\times$\\ 
$K^-\pi^+\pi^+\pi^-$ & $\times$ & $\times$\\ 
$K^-e^+\nu$ & $\times$ & \\ 
\hline
\multicolumn{3}{c}{CP-Even Tags} \\
\hline
$K^+K^-$  & $\times$ & $\times$\\ 
$\pi^+\pi^-$ & $\times$ & $\times$\\ 
$K_S^0\pi^0\pi^0$ & $\times$ & \\
$K_L^0\pi^0$ & $\times$ & \\ \hline
\multicolumn{3}{c}{CP-Odd Tags} \\
\hline
$K_S^0\pi^0$ & $\times$ & $\times$\\ 
$K_S^0\eta$ & $\times$ & $\times$\\ 
$K_S^0\omega$ & $\times$ & \\ \hline
$K_S^0\pi^+\pi^-$ & $\times$ & $\times$\\ 
\hline\hline
\label{table:modes}
\end{tabular}
\end{table}

\subsection{Single Tags}
The $\psi(3770)$ resonance is below threshold for $D\bar D \pi$ production, and so the events
of interest, $e^+e^-\to\psi(3770)\to D\bar D$, have $D$ mesons with energy equal to 
the beam energy and a unique momentum. Thus, for identifying $D^0$ candidates, we 
follow Mark III~\cite{MarkIII} and define two kinematic variables:
the beam-constrained candidate mass, $M_{BC}\equiv \sqrt{E_0^2/c^4-{\bf P}_D^2/c^2}$, where
${\bf P}_D$ is the $D^0$ candidate momentum and $E_0$ is the beam energy, and 
$\Delta E \equiv E_D-E_0$, where $E_D$ is the sum of the $D^0$ candidate daughter 
energies. Candidate tags are required to have $\Delta E$ within about 3 standard deviations of 
zero~\cite{TQCA}. 

For events with a $K^-\pi^+$, $K^+K^-$, and $\pi^+\pi^-$ single-tag (ST) that have no additional
charged particles, we apply additional selection requirements to suppress cosmic ray muons and 
Bhabha events. 
We do not
allow tracks identified as electrons or muons to be used in the tag.
We demand evidence of the other $D$ by requiring at least one electromagnetic shower in the 
calorimeter above 50 MeV not
associated with the tracks of the tag, where a single minimum ionizing particle deposits the equivalent
of 200 MeV.
For $K^+ K^-$ ST 
candidates, additional geometric requirements are needed to remove doubly radiative Bhabha events
followed by pair conversion of a radiated photon. We accept only one 
candidate per mode per event; when multiple candidates are present, we choose the one with
smallest $|\Delta E|$. 

\begin{figure}
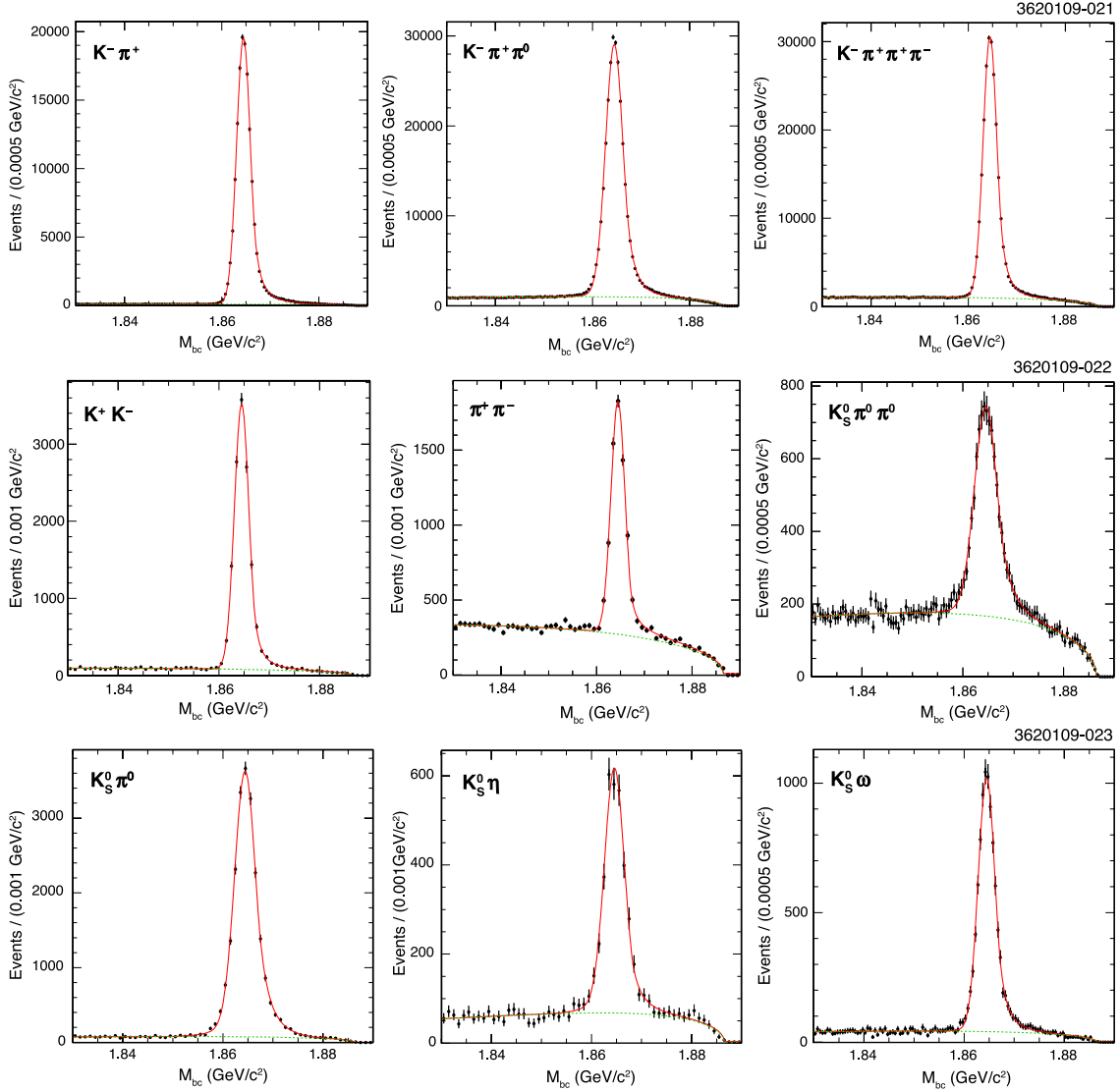

\centering
\includegraphics[width=15cm]{eps/fromJeanne/3620109-021.epsi}
\includegraphics[width=15cm]{eps/fromJeanne/3620109-022.epsi}
\includegraphics[width=15cm]{eps/fromJeanne/3620109-023.epsi}
\caption{Data $M_{bc}$ distribution for various tag modes.
The solid lines show the total fits, and the dashed lines show the background shapes.}
\label{fig:Mbc_Data}
\end{figure}

The resulting $M_{BC}$ distributions are shown in Fig.~\ref{fig:Mbc_Data}. Each distribution
is fit to a signal shape derived from simulated signal events and to a background
ARGUS~\cite{Argus} threshold function.
The ST yield is given by the area in the signal peak in the mass region from
$1.86<M_{BC}<1.87$~GeV.

\subsection{\boldmath Double tags with $K_S^0\pi^+\pi^-$}

We form $K_S^0\pi^+\pi^-$ DTs by combining a $K_S^0\pi^+\pi^-$ tag with a ST candidate.
We choose one DT candidate per mode per event with $\bar M$ closest to the measured $D^0$ mass,
where $\bar M \equiv [M(D^0)+M(\bar D^0)]/2$. 
 
Since the $D^0\to K_S^0 \pi^+\pi^-$ vs. $\bar D^0\to K_S^0\pi^+\pi^-$ sample plays a key
role in extracting $s_i$ values, 
we drop the requirement on the flight distance significance 
for $K_S^0$ candidates to increase the statistics. We find 421 
$D^0\to K_S^0 \pi^+\pi^-$ vs. $\bar D^0\to K_S^0\pi^+\pi^-$ candidates which include 
about $9\%$ background.  We increase the yield by about $15\%$ (additional 54 candidates, $\sim$15\% background) by reconstructing
the $K_S^0\pi^+\pi^-$ vs. $K_S^0\pi^+\pi^-$ candidates when one $\pi^{\pm}$ is 
not reconstructed. The presence of the $\pi^\pm$ is inferred from 
the missing four-momentum calculated from the well known initial state and the reconstructed particles.

\subsection{\boldmath Double tags with $K^-e^+\nu$}
\begin{figure}
\centering
\includegraphics[width=5cm]{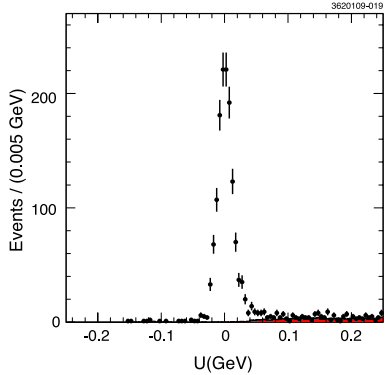}
\caption{$U\equiv E_{miss}-c|\vec{\mathbf p}_{miss}|$ distribution for $K^-e^+\nu$ in events with
a $D^0\to K_S^0\pi^+\pi^-$ signal candidate. The points are data and the shaded histogram represents 
a simulation of the backgrounds.}
\label{fig:Umiss}
\end{figure}

Candidate $K^-e^+\nu$ vs. $K_S^0\pi^+\pi^-$ DTs are reconstructed by combining a $K_S^0\pi^+\pi^-$
ST candidate with a kaon candidate and an electron candidate from the remainder of the 
event. Events with more than two additional tracks (aside from the $K^0_S\pi^+\pi^-$ daughters) are vetoed.
Signal discrimination for $D \to K^-e^+\nu$ uses the variable $U\equiv E_{miss}-c|\vec{\mathbf p}_{miss}|$,
where $E_{miss}$ and $\vec{\mathbf p}_{miss}$ are the missing energy and momentum in the semileptonic
$D^0$ meson decay, calculated using the difference of the four-momenta of the tag 
and that of the $K^-$ and $e^+$ candidates. For correctly identified events, $U=0$, since only 
the neutrino is undetected. After all selection criteria are applied, multiple candidates are rare for 
$K^-e^+\nu$. The $U$ distribution for $D^0\to K^-e^+\nu$ candidates is shown in Fig.~\ref{fig:Umiss}. 
The points with error bars are data and the shaded histogram represents a simulation of the background, 
which is less than $1\%$ in the signal region, $|U|<50$ MeV.

\subsection{\boldmath Double tags with $K_L^0 \pi^0$}
\label{section:KlPi0}

\begin{figure}
\centering
\includegraphics[width=5cm]{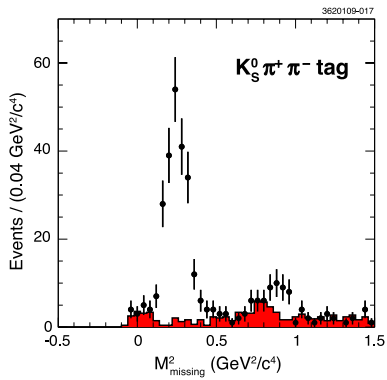}
\caption{$M_{miss}^2$ distributions for $K_L^0\pi^0$ when one $D$ is identified
as $K_S^0\pi^+\pi^-$. Shaded histogram represents simulation of the backgrounds.
The enhancement of data relative to simulation $\sim$ $0.9~{\rm GeV}^2/c^4$ corresponds to the decay
$K^{*0} \pi^0 \to K^0_L\pi^0 \pi^0$ where both the $K^0_L$ and $\pi^0$ from the $K^{*0}$ are undetected. The Dalitz-plot model of this process is not implemented in our simulation.}
\label{fig:MM2_KlPi0}
\end{figure}

The $K_S^0\pi^+\pi^-$ vs. $K_L^0\pi^0$ DT mode is reconstructed with a missing mass technique
since the $K_L^0$ mesons produced at CLEO-c are not reconstructed. A fully reconstructed 
$K_S^0\pi^+\pi^-$ ST is combined with a $\pi^0$ candidate, and we compute the 
recoil-mass squared against the ST-$\pi^0$ system, $M_{miss}^2$. Signal $K_L^0\pi^0$ decays 
are identified by a peak in $M_{miss}^2$ at $M^2_{K_L^0}$. Backgrounds from $D \to K_S^0 \pi^0$,
$\pi^0\pi^0$, and $\eta \pi^0$ are suppressed by vetoing events with additional unassigned
charged particles, or $\eta\to\gamma\gamma$ or $\pi^0\to\gamma\gamma$ candidates.
We further suppress backgrounds by making requirements on 
the energy of showers in the calorimeter that are not associated with the decay products of the
$K_S^0\pi^+\pi^-$ or the $\pi^0$. 
We compute the angle, $\theta$, between each unassigned shower 
and the direction of the missing momentum. For 
$\cos\theta<0.9$, we require the energy of showers, $E_{shower} <100$ MeV for any single shower. If 
$0.9<\cos\theta<0.98$, we require $E_{shower}<100+250\times (\cos\theta-0.9)$ MeV.
The $M_{miss}^2$ distribution for $K_L^0 \pi^0$ is shown in Fig.~\ref{fig:MM2_KlPi0}.
The points with error bars are data and the shaded histogram represents a simulation of the backgrounds.
Signal candidates are required to be within the range $0.1<M_{miss}^2<0.5$~GeV$^2/c^4$.

\subsection{\boldmath Double tags with $K_L^0 \pi^+\pi^-$}
\begin{figure}
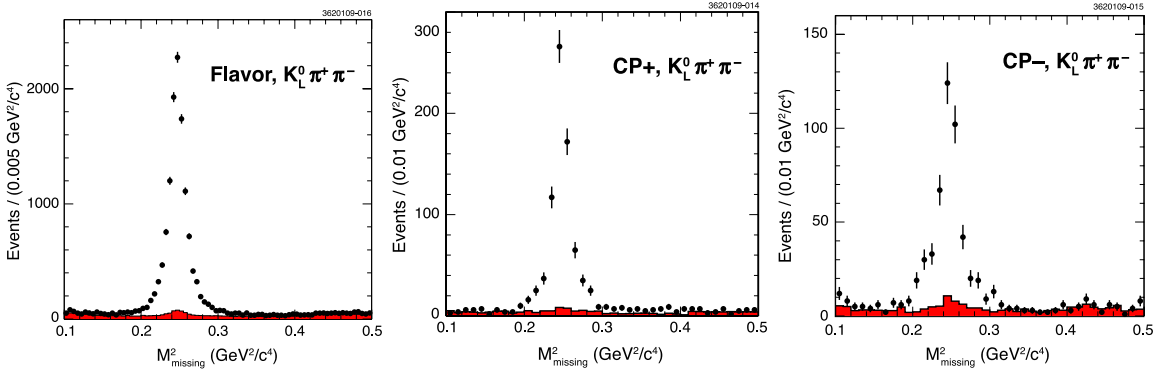

\centering
\includegraphics[width=5cm]{eps/fromJeanne/3620109-016.epsi}
\includegraphics[width=5cm]{eps/fromJeanne/3620109-014.epsi}
\includegraphics[width=5cm]{eps/fromJeanne/3620109-015.epsi}
\caption{$M_{miss}^2$ distributions for $K_L^0\pi^+\pi^-$ for flavor tags (left),
CP-even tags (middle), and CP-odd tags (right). The points with error bars are data and 
the shaded histograms represent simulations of the backgrounds.}
\label{fig:MM2_KlPP}
\end{figure}

Candidate $K_L^0 \pi^+\pi^-$ decays are reconstructed in DTs using a similar missing mass 
technique as described in Section~\ref{section:KlPi0}. We require the signal side 
(associated with the $K_L^0 \pi^+\pi^-$ candidate) to have exactly two charged tracks. 
Backgrounds are reduced by applying $\pi^0$, $\eta$, and $K_S^0$ vetoes. 
Using the measured momenta of the tagged $D^0$ and the two additional pions, 
we compute the missing momentum and missing energy on the signal side. 
We apply the same requirements to the energy of the unassigned showers as described in 
Section~\ref{section:KlPi0}.
The $M_{miss}^2$ distributions for $K_L^0 \pi^+\pi^-$ are shown in Fig.~\ref{fig:MM2_KlPP}.
The points with uncertainties are data and the shaded histograms show a simulation of the backgrounds.
Signal events are required to have a missing-mass squared in the region
$0.21<M_{miss}^2<0.29$ GeV$^2/c^4$.

\subsection{Yields in Data}
\begin{figure}
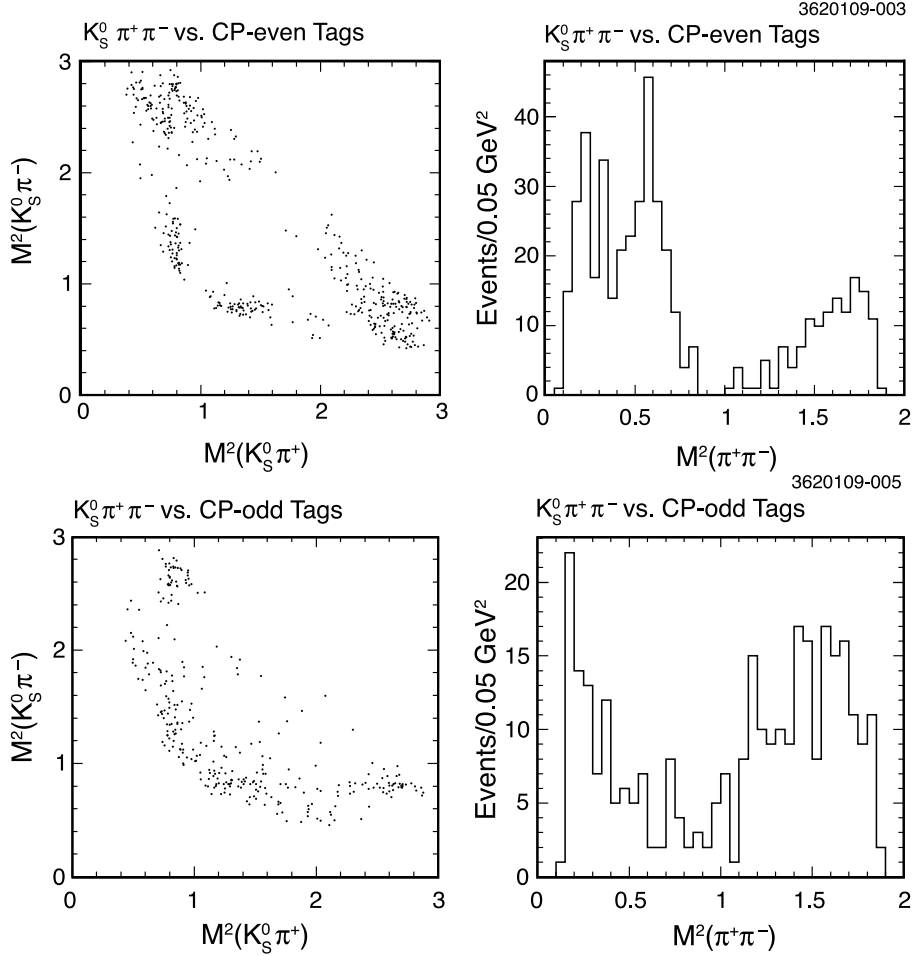

\centering
\includegraphics[width=12cm]{eps/fromJeanne/3620109-003.epsi}
\includegraphics[width=12cm]{eps/fromJeanne/3620109-005.epsi}
\caption{CP-even tagged $K_S^0\pi^+\pi^-$ Dalitz plot (a), and 
its $m^2(\pi^+\pi^-)$ projection (b). 
CP-odd tagged $K_S^0\pi^+\pi^-$ Dalitz plot (c), and 
its $m^2(\pi^+\pi^-)$ projection (d).}
\label{fig:CP_KsPP}
\end{figure}

The ST yields for the tag modes and DT yields for $K_{S/L}^0 \pi^+\pi^-$ versus 
different tags are shown in Table~\ref{table:yields}. To determine the
$D^0\to K^-e^+\nu$ and $D^0\to K_L^0\pi^0$ ST yields
we use the integrated luminosity, measured $D^0\bar{D}^0$ cross-sections~\cite{DHadronic}
and measured branching fractions~\cite{CLEO-SL,Qing}. Combining all modes of the
same CP, we show in Fig.~\ref{fig:CP_KsPP} the Dalitz-plot distribution of 
CP-even and CP-odd tagged $\tilde D^0\to K_S^0\pi^+\pi^-$ decays. 
Figure~\ref{fig:CP_KlPP} shows the corresponding distributions for
CP-tagged $\tilde D^0\to K_L^0\pi^+\pi^-$ decays. The clear absence of
a $\rho^0 K^0_S$ component (CP-odd) in CP-odd tagged $K^0_S\pi^+\pi^-$ decays
is an illustration of the quantum correlations that exist in the 
$\psi(3770)\to D^0\bar{D}^0$ decay. For $K^0_L\pi^+\pi^-$, $\rho^0 K_L^0$ is 
absent in the CP-even tagged samples.

The signal-to-background ratios in our $K_{S/L}^0\pi^+\pi^-$ DT samples range from
10 to better than 100, depending on tag mode. The tag side $\Delta E$, $K_S^0$
and $\omega$ sidebands are used for combinatorial and non-resonant background subtraction. 
On the signal side, the background level is 1.9\% for $K_S^0\pi^+\pi^-$ after applying
the $K_S^0$ flight significance requirement. This part of the background is 
considered as a systematic error.
The background-to-signal ratio for the $K_L^0\pi^+\pi^-$ signal side is about 5$\%$,
of which about 2\% is a peaking background from $K_S^0\pi^+\pi^-,~K_S^0\to\pi^0\pi^0$ 
decays that pass the $K_L^0\pi^+\pi^-$ selection criteria.
We estimate this peaking background yield using $K_S^0\pi^+\pi^-$ data 
and a misidentification rate determined from a quantum-correlated Monte Carlo simulation. 
The combinatorial background contribution is estimated using the $M_{miss}^2$ sidebands.
The expected yields from these two background sources are subtracted from the observed
signal yields to obtain background-corrected yields.

For the $K_S^0\pi^+\pi^-$ vs. $K_S^0\pi^+\pi^-$ sample, no $K_S^0$ flight significance requirement 
was applied, resulting in a background-to-signal ratio of $\sim$9\%. About 7\% (of 9\%) of this background comes from $\tilde D^0\to\pi^+ \pi^- \pi^+ \pi^-$ faking $K_S^0 \pi^+ \pi^-$.
This background is subtracted using a Monte Carlo simulation of this decay, where
the $\pi^+ \pi^- \pi^+ \pi^-$ Dalitz-plot structure is taken from the FOCUS experiment~\cite{PPPP}.
The impact of the remaining $\sim$1.9\% of background on our nominal fit results is small and included in the systematic uncertainties.

\begin{table}[t]
\centering
\caption{Single tag and $K_{S/L}^0\pi^+\pi^-$ double tag yields.}
\begin{tabular}{lccc}        \hline\hline
 Mode  & ST Yield & $K_S^0\pi^+\pi^-$ yield & $K_L^0\pi^+\pi^-$ yield\\\hline
\multicolumn{4}{c}{Flavor Tags} \\
\hline
$K^-\pi^+$ & 144563 $\pm$ 403 &1447 & 2858 \\
$K^-\pi^+\pi^0$  & 258938 $\pm$ 581 & 2776 & 5130\\
$K^-\pi^+\pi^+\pi^-$  & 220831 $\pm$ 541 & 2250 & 4110\\
$K^-e^+\nu$  &  123412 $\pm$ 4591& 1356 & -\\
\hline
\multicolumn{4}{c}{CP-Even Tags} \\
\hline
$K^+K^-$  & 12867 $\pm$ 126 & 124 & 345 \\
$\pi^+\pi^-$  & 5950 $\pm$ 112 & 62 & 172\\
$K_S^0\pi^0\pi^0$   &6562 $\pm$ 131 & 56 & -\\
$K_L^0\pi^0$ & 27955 $\pm$ 2013 & 229 & -\\
\hline
\multicolumn{4}{c}{CP-Odd Tags} \\
\hline
$K_S^0\pi^0$  & 19059 $\pm$ 150 & 189 & 281\\ 
$K_S^0\eta$  & 2793 $\pm$ 69 & 39 & 41\\
$K_S^0\omega$  & 8512 $\pm$ 107 & 83 & -\\ 
\hline
$K_S^0\pi^+\pi^-$  & -& 475  & 867\\
\hline\hline
\label{table:yields}
\end{tabular}
\end{table}

The reconstruction efficiency is defined as the ratio of reconstructed events to 
generated events in each bin.
The reconstruction efficiencies are calculated from large Monte Carlo samples 
generated according to the amplitude description of Eqs.~\ref{eqn:Amp_CP_Ks} and~\ref{eqn:Amp_Ks_Ks} 
for different tag modes. Dividing the observed yields
in each $\delta_D$ bin by this efficiency, 
we obtain the efficiency-corrected yields, $M_i^{\pm}$ and $M_{ij}$.

\begin{figure}
\centering
\includegraphics[width=12cm]{eps/fromJeanne/3620109-002.epsi}
\includegraphics[width=12cm]{eps/fromJeanne/3620109-004.epsi}
\caption{CP-even tagged $K_L^0\pi^+\pi^-$ Dalitz plot (a), and 
its $m^2(\pi^+\pi^-)$ projection (b). 
CP-odd tagged $K_L^0\pi^+\pi^-$ Dalitz plot (c), and 
its $m^2(\pi^+\pi^-)$ projection (d).}
\label{fig:CP_KlPP}
\end{figure}

\section{\boldmath Extraction of $c_i$ and $s_i$}
\label{section:fit}
We determine the coefficients $c_i$, $s_i$ by minimizing the negative log-likelihood function

\begin{eqnarray}
-2{\rm log}\mathcal L &=& -2\sum_i{\rm log}P(M_i^{\pm}, \langle M_i^{\pm} \rangle)_{(CP, K_S^0\pi^+\pi^-)}\nonumber\\
&& -2\sum_i{\rm log}P(M_i^{\pm}, \langle M_i^{\pm} \rangle)_{(CP, K_L^0\pi^+\pi^-)} \nonumber\\
&& -2\sum_{i,j}{\rm log}P(M_{ij}, \langle M_{ij} \rangle)_{(K_S^0\pi^+\pi^-, K_S^0\pi^+\pi^-)} \nonumber\\
&& - 2\sum_{i,j}{\rm log}P(M_{ij}, \langle M_{ij} \rangle)_{(K_S^0\pi^+\pi^-, K_L^0\pi^+\pi^-)}\nonumber\\
&& + \chi^2, 
\label{eqn:logL}
\end{eqnarray}

\noindent where $\langle M_i^{\pm} \rangle$ is calculated according Eqs.~\ref{eqn:CP_Ks} and 
~\ref{eqn:CP_Kl}, and $\langle M_{ij}\rangle$ is calculated according 
Eqs.~\ref{eqn:Ks_Ks} and~\ref{eqn:Ks_Kl}, and
$P(M,\langle M \rangle)$ is the Poisson probability to get $M$ events given the
expected number, $\langle M \rangle$. In our nominal fit, a $\chi^2$ penalty term 
\begin{equation}
\chi^2 = \sum_i(\frac{c_i'-c_i-\Delta c_i}{\delta\Delta c_i})^2+
\sum_i(\frac{s_i'-s_i-\Delta s_i}{\delta\Delta s_i})^2,
\end{equation}
constrains $c_i'$ and
$s_i'$ to differ from $c_i$ and $s_i$, respectively, by their expected differences $\Delta c_i$, 
$\Delta s_i$, within errors. (Those errors, $\delta\Delta c_i, \delta\Delta s_i$, are the systematic 
uncertainties shown in Table~\ref{table:correction}.) This constraint has little impact on $c_i$ but is important for $s_i$ and will be relaxed and tightened as a systematic variation.

From Monte Carlo studies, we found that DCSD decays in 
flavor tag modes ($K^- \pi^+$, $K^- \pi^+\pi^0$, $K^- \pi^+\pi^+\pi^-$) lead to a
significant bias in the $K_{i}^{(\prime)}$'s due to an interference of the wrong flavor of the 
$\tilde{D}^{0}\to K^{0}_{S/L}\pi^{+}\pi^{-}$ decay; this results in a significant bias in the values of $c_i^{(\prime)}$ and $s_i^{(\prime)}$ from the $D^0 \to K^0_{S}\pi^+\pi^-$ vs. $D^0\to K^0_{S/L}\pi^+\pi^-$  analyses.
Therefore, for the 
$D^0\to K^0_{L}\pi^+\pi^-$ vs. $D^0\to K^0_{S}\pi^+\pi^-$ analysis, we use only the
$D^0\to K^-e^+\nu$ tagged $K_S^0\pi^+\pi^-$ sample and the 
$D^0\to K^-\pi^+$ tagged $K_L^0\pi^+\pi^-$ 
sample for counting $S_f^{(')}$ and $K_{i}^{(')}$ yields.\footnote{All three hadronic flavor tag modes are used to determine $K_i^{(\prime)}$ for the CP-tag vs. $K^0_{S/L} \pi^+\pi^-$ determination of $c_i^{(\prime)}$.}
For the latter, we estimate the biases and adjust the $K_i^{(')}$ values using the correction factor:

\begin{displaymath}
|A_{D^0 \to K_S^0\pi^+\pi^-}|^2/|A_{D^0\to K_S^0\pi^+\pi^-}+
re^{-i\delta}A_{\bar D^0\to K_S^0\pi^+\pi^-}|^2. 
\end{displaymath}

\noindent Here $r=|A(D^0\to K^+\pi^-)/A(D^0\to K^-\pi^+)|$ and $\delta_{K\pi}$
are the ratio of amplitudes of the DCSD to CF decay and the relative strong phase, respectively.
The amplitude ratio squared, $r^2=(3.44\pm0.01\pm0.09)\times10^{-3}$ and $\delta_{K\pi}=(22\pm16.3)^\circ$ 
are taken from Ref.~\cite{TQCA}. This correction factor is estimated in each of our eight
Dalitz-plot bins using the BaBar $D^0\to K_S^0 \pi^+ \pi^-$ Dalitz-plot fit amplitude~\cite{GGSZ_BaBar1}.
The model dependence of this correction is negligible. Uncertainties on these corrections
due to the uncertainty on $\delta_{K\pi}$ are
small and are included in our systematic uncertainties.

The fitting procedure was tested using a simulated $C$-odd $D^0\bar D^0$ Monte Carlo sample
where we performed 100 toy $K^0_S\pi^+\pi^-$ vs. $K^0_S\pi^+\pi^-$ experiments
with $c_i$ and $s_i$ taken from the BaBar model. The means and widths of the
pull distributions of the $c_i$ and $s_i$ parameters were consistent with zero
and one, respectively, indicating no bias and proper estimation of statistical uncertainties.

To enable the separation of  the statistical uncertainty on $c_i$ and $s_i$ from the systematic uncertainty on $\Delta c_i$ and $\Delta s_i$ we perform a likelihood fit to $(c_i,s_i)$ with the values of 
$(c_i', s_i')$ fixed according to Table~\ref{table:correction}. The results of this fit and of the nominal likelihood fit to $(c_i, s_i)$, $(c_i', s_i')$ 
are shown in Table~\ref{table:ci_si}. The (statistical) correlation matrix among $c_i$ and $s_i$ 
in the constrained fit is
shown in Table~\ref{table:correlation}. Note that in Table~\ref{table:ci_si}, we choose the 
$s_i$ and $s_i'$ signs based on BaBar isobar model predictions~\cite{GGSZ_BaBar1} 
to resolve the two-fold ambiguity discussed in Section~\ref{section:diff_KsKl}.

\begin{table}[t]
\centering
\caption{Fit results for $c_i$, $s_i$ (with $\Delta c_i$ and $\Delta s_i$ fixed), $c_i$, $c_i'$, $s_i$ and $s_i'$ (with $\Delta c_i$ and $\Delta s_i$ constrained).  See Table~\ref{table:correction} for $\Delta c_i$ and $\Delta s_i$.} 
\begin{tabular}{ccc|cccc}        \hline\hline
 & \multicolumn{2}{c}{$\Delta c_i$, $\Delta s_i$ fixed} &\multicolumn{4}{c}{$\Delta c_i$,$\Delta s_i$ constrained}\\
  $i$ &  $c_i$ & $s_i$   & $c_i$ & $c_i'$  & $s_i$ & $s_i'$ \\\hline
  0 & $\hphantom{-}0.742  \pm  0.037$ & $  0.004  \pm  0.160$ & $\hphantom{-}0.743 \pm 0.041$ & $\hphantom{-}0.840 \pm 0.041$ & $\hphantom{-}0.014 \pm 0.166$ & $-0.021 \pm 0.164$ \\
 1 & $\hphantom{-}0.606  \pm  0.071  $ & $\hphantom{-}0.014  \pm  0.215$ & $\hphantom{-}0.611 \pm 0.072$ & $\hphantom{-}0.779 \pm 0.072$ & $\hphantom{-}0.014 \pm 0.216$ & $-0.069 \pm 0.219$ \\ 
 2 & $ -0.008  \pm  0.063$ & $\hphantom{-}0.581  \pm  0.190$ & $\hphantom{-}0.059 \pm 0.077$ & $\hphantom{-}0.250 \pm 0.078$ & $\hphantom{-}0.609 \pm 0.188$ & $\hphantom{-}0.587 \pm 0.188$ \\ 
 3 & $ -0.529  \pm  0.101$ & $\hphantom{-}0.138  \pm  0.217$ & $-0.495 \pm 0.114$ & $-0.349 \pm 0.135$ & $\hphantom{-}0.151 \pm 0.225$ & $\hphantom{-}0.275 \pm 0.232$ \\ 
 4 & $ -0.889  \pm  0.049$ & $-0.053  \pm  0.183$ & $-0.911 \pm 0.053$ & $-0.793 \pm 0.057$ & $-0.050 \pm 0.189$ & $-0.016 \pm 0.192$ \\ 
 5 & $ -0.742  \pm  0.066$ & $-0.317  \pm  0.187$ & $-0.736 \pm 0.070$ & $-0.546 \pm 0.080$ & $-0.340 \pm 0.194$ & $-0.388 \pm 0.200$ \\ 
 6 & $\hphantom{-}0.108  \pm  0.074$ & $ -0.836  \pm  0.185$ & $\hphantom{-}0.157 \pm 0.092$ & $\hphantom{-}0.475 \pm 0.094$ & $-0.827 \pm 0.190$ & $-0.725 \pm 0.196$ \\ 
 7 & $\hphantom{-}0.403  \pm  0.046$ & $ -0.410  \pm  0.158$ & $\hphantom{-}0.403 \pm 0.046$ & $\hphantom{-}0.591 \pm 0.048$ & $-0.409 \pm 0.158$ & $-0.374 \pm 0.169$ \\
\hline\hline
\label{table:ci_si}
\end{tabular}
\end{table}

\begin{table*}[hbt]
\centering
\scriptsize
\caption{Correlation Matrix for the $c_i$ and $s_i$ parameters. Labels 1-8 represents $c_1-c_8$ 
and 9-16 represent $s_1-s_8$.}
\begin{tabular}{r|rrrrrrrrrrrrrrrr}        \hline\hline
$i$   & 1&2 &3 &4 &5 &6 & 7 & 8 & 9&10 &11 &12 &13 &14 & 15 & 16 \\\hline
 1 &  1.000& & & & & & & & & & & & & & & \\
 2 & -0.028&  1.000& & & & & & & & & & & & & & \\
 3 & -0.007& -0.011&  1.000& & & & & & & & & & & & & \\
 4 &  0.035&  0.011&  0.002&  1.000& & & & & & & & & & & & \\
 5 &  0.073& -0.022&  0.006&  0.004&  1.000& & & & & & & & & & & \\
 6 &  0.016&  0.069& -0.003&  0.003& -0.104&  1.000& & & & & & & & & & \\
 7 &  0.018&  0.013& -0.020&  0.005&  0.033&  0.016&  1.000& & & & & & & & & \\
 8 & -0.020& -0.028&  0.008&  0.005&  0.050&  0.013&  0.015&  1.000& & & & & & & & \\
 9 &  0.024&  0.006& -0.072& -0.006&  0.014&  0.014&  0.113&  0.040&  1.000& & & & & & & \\
10 &  0.000& -0.033&  0.017& -0.007& -0.003&  0.038& -0.001& -0.001& -0.060&  1.000& & & & & & \\
11 &  0.006&  0.007& -0.025& -0.002&  0.003& -0.008&  0.041&  0.029&  0.323& -0.154&  1.000& & & & & \\
12 &  0.004&  0.005& -0.020&  0.035&  0.002&  0.000&  0.014&  0.005&  0.149& -0.124&  0.244&  1.000& & & & \\
13 &  0.001& -0.001& -0.003& -0.014& -0.008& -0.072& -0.007& -0.004&  0.158& -0.107&  0.340&  0.070&  1.000& & & \\
14 & -0.011& -0.014&  0.078&  0.005& -0.004& -0.042& -0.086& -0.021& -0.448&  0.085& -0.213& -0.124& -0.275&  1.000& & \\
15 &  0.009&  0.008& -0.053& -0.003&  0.004&  0.002&  0.061&  0.013&  0.373& -0.139&  0.314&  0.228&  0.269& -0.405&  1.000& \\
16 &  0.004&  0.004& -0.056& -0.004&  0.003& -0.007&  0.026&  0.041&  0.234& -0.096&  0.521&  0.176&  0.243& -0.133&  0.106&  1.000\\
\hline\hline
\end{tabular}
\label{table:correlation}
\end{table*}

\section{Systematic uncertainties}
Systematic uncertainties on ($c_i$, $s_i$) and ($c'_i$, $s'_i$) come from many sources.
Table~\ref{table:sys_ci} and Table~\ref{table:sys_si} summarize the 
main contributions of the systematic uncertainties for  
$c_i$ and $s_i$, respectively. 
Table~\ref{table:sys_c'i} and Table~\ref{table:sys_s'i} summarize the 
main contributions of the systematic uncertainties for  
$c'_i$ and $s'_i$, respectively. 
\begin{table}[h]
\centering
\footnotesize
\caption{ Systematic uncertainties for $c_i$.}
\begin{tabular}{lrrrrrrrr}        \hline\hline
 & $c_1$ & $c_2$ & $c_3$ & $c_4$ & $c_5$ & $c_6$ & $c_7$ & $c_8$
 \\\hline
$K_i^{(')}$ statistics error & 0.010 & 0.015 & 0.016 & 0.019 & 0.009 & 0.015 & 0.018 & 0.008\\
Momentum resolution & 0.008 & 0.015 & 0.012 & 0.019 & 0.011 & 0.012 & 0.017 & 0.009\\
Efficiency variation & 0.004 & 0.007 & 0.011 & 0.008 & 0.005 & 0.008 & 0.010 & 0.006\\
Single Tag yields & 0.006 & 0.007 & 0.013 & 0.011 & 0.005 & 0.008 & 0.015 & 0.008\\
Tag side background & 0.007 & 0.007 & 0.014 & 0.013 & 0.006 & 0.008 & 0.014 & 0.011\\
$K_S^0\pi^+\pi^-$ background & 0.001 & 0.002 & 0.009 & 0.027 & 0.012 & 0.006 & 0.003 & 0.002\\
$K_L^0\pi^+\pi^-$ background & 0.006 & 0.018 & 0.004 & 0.024 & 0.017 & 0.012 & 0.020 & 0.006\\
Multi-Candidate selection & 0.002 & 0.003 & 0.003 & 0.008 & 0.004 & 0.006 & 0.003 & 0.002\\
Non-$D/\bar D$ & 0.010 & 0.016 & 0.004 & 0.007 & 0.004 & 0.005 & 0.006 & 0.005\\
DCSD & 0.009 & 0.012 & 0.005 & 0.013 & 0.014 & 0.010 & 0.013 & 0.006\\\hline
Sum & 0.022 & 0.037 & 0.031&  0.052 & 0.032 & 0.030 & 0.042 & 0.021 \\
\hline\hline
\end{tabular}
\label{table:sys_ci}

\centering
\footnotesize
\caption{ Systematic uncertainties for $s_i$.}
\begin{tabular}{lrrrrrrrr}        \hline\hline
 & $s_1$ & $s_2$ & $s_3$ & $s_4$ & $s_5$ & $s_6$ & $s_7$ & $s_8$
 \\\hline
$K_i^{(')}$ statistics error & 0.031 & 0.027 & 0.039 & 0.030 & 0.023 & 0.023 & 0.033 & 0.026\\
Momentum resolution & 0.018 & 0.035 & 0.023 & 0.033 & 0.023 & 0.022 & 0.022 & 0.018\\
Efficiency variation & 0.018 & 0.012 & 0.019 & 0.010 & 0.013 & 0.018 & 0.013 & 0.012\\
Single Tag yields & 0.005 & 0.001 & 0.005 & 0.004 & 0.003 & 0.003 & 0.005 & 0.003\\
Tag side background & 0.004 & 0.001 & 0.001 & 0.003 & 0.001 & 0.004 & 0.002 & 0.001\\
$K_S^0\pi^+\pi^-$ background & 0.005 & 0.008 & 0.030 & 0.023 & 0.005 & 0.003 & 0.016 & 0.014\\
$K_L^0\pi^+\pi^-$ background & 0.050 & 0.022 & 0.018 & 0.035 & 0.006 & 0.024 & 0.005 & 0.025\\
Multi-Candidate selection & 0.036 & 0.018 & 0.033 & 0.012 & 0.022 & 0.026 & 0.028 & 0.011\\
Non-$D/\bar D$ & 0.005 & 0.003 & 0.004 & 0.005 & 0.005 & 0.005 & 0.003 & 0.002\\
DCSD & 0.023 & 0.004 & 0.030 & 0.019 & 0.015 & 0.006 & 0.027 & 0.020\\\hline
Sum & 0.077 & 0.055 & 0.076 & 0.069 & 0.045 & 0.052 & 0.060 & 0.050 \\
\hline\hline
\end{tabular}
\label{table:sys_si}
\end{table}

\begin{table}[t]
\centering
\footnotesize
\caption{ Systematic uncertainties for $c'_i$.}
\begin{tabular}{lrrrrrrrr}        \hline\hline
 & $c_1$ & $c_2$ & $c_3$ & $c_4$ & $c_5$ & $c_6$ & $c_7$ & $c_8$
 \\\hline
$K_i^{(')}$ statistics error & 0.009 & 0.014 & 0.011 & 0.013 & 0.009 & 0.012 & 0.017 & 0.008\\
Momentum resolution & 0.008 & 0.016 & 0.014 & 0.020 & 0.010 & 0.013 & 0.019 & 0.010\\
Efficiency variation & 0.006 & 0.006 & 0.009 & 0.009 & 0.004 & 0.007 & 0.011 & 0.006\\
Single Tag yields & 0.005 & 0.007 & 0.007 & 0.007 & 0.003 & 0.005 & 0.006 & 0.007\\
Tag side background & 0.005 & 0.007 & 0.007 & 0.007 & 0.004 & 0.005 & 0.005 & 0.009\\
$K_S^0\pi^+\pi^-$ background & 0.002 & 0.001 & 0.005 & 0.012 & 0.007 & 0.003 & 0.001 & 0.002\\
$K_L^0\pi^+\pi^-$ background & 0.008 & 0.020 & 0.017 & 0.046 & 0.019 & 0.014 & 0.015 & 0.006\\
Multi-Candidate selection & 0.003 & 0.003 & 0.001 & 0.004 & 0.003 & 0.004 & 0.002 & 0.002\\
Non-$D/\bar D$ & 0.012 & 0.018 & 0.005 & 0.006 & 0.004 & 0.004 & 0.008 & 0.005\\
DCSD & 0.011 & 0.013 & 0.007 & 0.013 & 0.014 & 0.013 & 0.014 & 0.006\\ \hline
Sum & 0.023 & 0.039 & 0.029 & 0.057 & 0.029 & 0.028 & 0.036 & 0.021 \\
\hline\hline
\end{tabular}
\label{table:sys_c'i}

\centering
\footnotesize
\caption{ Systematic uncertainties for $s'_i$.}
\begin{tabular}{lrrrrrrrr}        \hline\hline
 & $c_1$ & $c_2$ & $c_3$ & $c_4$ & $c_5$ & $c_6$ & $c_7$ & $c_8$
 \\\hline
$K_i^{(')}$ statistics error & 0.030 & 0.028 & 0.037 & 0.026 & 0.022 & 0.025 & 0.034 & 0.028\\
Momentum resolution & 0.022 & 0.042 & 0.031 & 0.042 & 0.029 & 0.028 & 0.031 & 0.033\\
Efficiency variation & 0.018 & 0.012 & 0.015 & 0.009 & 0.012 & 0.018 & 0.013 & 0.013\\
Single Tag yields & 0.005 & 0.001 & 0.005 & 0.003 & 0.004 & 0.003 & 0.004 & 0.003\\
Tag side background & 0.003 & 0.000 & 0.002 & 0.002 & 0.002 & 0.004 & 0.003 & 0.001\\
$K_S^0\pi^+\pi^-$ background & 0.006 & 0.009 & 0.028 & 0.021 & 0.004 & 0.003 & 0.015 & 0.015\\
$K_L^0\pi^+\pi^-$ background & 0.052 & 0.022 & 0.013 & 0.033 & 0.003 & 0.025 & 0.004 & 0.025\\
Multi-Candidate selection & 0.038 & 0.018 & 0.031 & 0.007 & 0.020 & 0.027 & 0.028 & 0.013\\
$\delta\Delta c_i$, $\delta\Delta s_i$ & 0.034 & 0.048 & 0.001 & 0.006 & 0.009 & 0.072 & 0.050 & 0.048\\
Non-$D/\bar D$ & 0.005 & 0.003 & 0.004 & 0.005 & 0.004 & 0.005 & 0.003 & 0.002\\
DCSD & 0.021 & 0.005 & 0.028 & 0.016 & 0.013 & 0.007 & 0.028 & 0.022\\ \hline
Sum & 0.080 & 0.060 & 0.072 & 0.067 & 0.046 & 0.056 & 0.065 & 0.059 \\
\hline\hline
\end{tabular}
\label{table:sys_s'i}
\end{table}

In the global fit, the fitter does not take the statistical uncertainties associated with flavor 
tagged samples into account.
We estimate this part of the uncertainties by varying the input variables ($K_i^{(')}$) 
one by one according to their
statistical uncertainties, and by making new fits. At the end, we take the quadratic sum of all the 
variations as the systematic error.

Since our Dalitz-plot binning results in bins with unusual shapes and in some cases very narrow regions (see Fig.~\ref{fig:phasebin}), the migration of events from one bin
to another bin may bias our result. The position of an event in the Dalitz plot depends on its momentum 
determination. The systematic error associated with momentum resolution is studied by smearing the 
momentum of a fully simulated Monte Carlo 200 times,
according to the CLEO detector momentum resolution.
The distributions of the results for ($c_i$, $s_i$) and ($c'_i$, $s'_i$) are then fitted with Gaussian functions, and the widths of the distributions are taken as the systematic uncertainties. 

The systematic uncertainties associated with $K_{S/L}^0 \pi^+\pi^-$ finding cancel 
under the assumption that the efficiency systematic uncertainties are uniform across the Dalitz plot. Under this condition Eq.~\ref{eqn:CP_Ks}, Eq.~\ref{eqn:Ks_Ks}, Eq.~\ref{eqn:CP_Kl}, and Eq.~\ref{eqn:Ks_Kl}
have the same dependence on efficiency.
To account for a small non-uniformity, we generate a large number of toy experiments where we randomly distribute the efficiency of each bin according to a Gaussian distribution (width is taken as 0.02) 
and repeat this process for many times. The widths of the resulting distributions for ($c_i$, $s_i$) and ($c'_i$, $s'_i$) are taken as systematic uncertainties. 

The systematic uncertainties associated with the single tag yields are evaulated by repeating 
the fit with the input values varied by their own uncertainties.
Assuming the contributions are uncorrelated, we sum in quadrature to obtain the uncertainties 
on ($c_i$, $s_i$) and ($c'_i$, $s'_i$) due to single tag yields given
in Table~\ref{table:sys_ci} - Table~\ref{table:sys_s'i}.

The systematic uncertainties due to the estimation of the tag side background are studied mode by mode, and the quadratic sum is given in Table~\ref{table:sys_ci} - Table~\ref{table:sys_s'i}.

Though we used $\Delta E$ and $M(\pi^+\pi^-)$ mass sidebands for the tag side backgrounds 
subtraction, we did not apply any background subtraction for $K_S^0 \pi^+ \pi^-$ 
signal side, which is believed to be small since we require the decay vertex of $K_S^0$ to be
separated from the interaction region with a significance greater than two standard deviations.
The background level in the signal region is estimated from
$M(\pi^+\pi^-)$ sidebands. We found there is about a $1.9\%$ background in the signal region.
The systematic uncertainties due to this part of the background are estimated using Quantum Correlated 
Monte Carlo samples. We estimate the background contributions from Quantum Correlated Monte 
Carlo samples, then make a new fit with the background subtracted. The differences of the results 
between the nominal fit and the new fit are taken as the systematic uncertainties. 

The systematic uncertainties due to the $K_L^0\pi^+\pi^-$ background shape are considered by repeating
the fit assuming the background across the Dalitz plot is uniform. 
The uncertainties due to the estimation of the background
level are negligible.
The systematic uncertainties due to flavor-tagged,
CP-even tagged, CP-odd tagged, and $K_S^0 \pi^+ \pi^-$ tagged $K_L^0 \pi^+ \pi^-$ samples
are considered separately and summed in quadrature in 
Table~\ref{table:sys_ci} - Table~\ref{table:sys_s'i}.

It is possible to select a wrong  combination when there are multiple signal candidates 
in an event, especially for $K_S^0\pi^+\pi^-$ vs. $K_{S/L}^0\pi^+\pi^-$ samples, since
there are many pions with similar momenta. The
systematic uncertainties are studied by applying
correction matrices to the yield matrices $M_{ij}$. The corrections are typically 2\% (5\%) for 
the $K_S^0\pi^+\pi^-$ vs. $K_{S}^0\pi^+\pi^-$ ($K_S^0\pi^+\pi^-$ vs. $K_{L}^0\pi^+\pi^-$) event samples.

Monte Carlo simulated continuum events are checked for non-$D^0/\bar D^0$ backgrounds. No significant peaking
background is seen for double tagged $K_S^0 \pi^+ \pi^-$ samples. The contributions for 
$K^+ K^-$, $\pi^+\pi^-$, and $K_S^0\eta$ tagged $K_L^0 \pi^+ \pi^-$ samples are also negligible.
For other samples, there are 1$\sim$2\% contributions depending on the tag mode. 
A systematic study is performed by
assuming the background is uniformly distributed over the Dalitz plot.

For the DCSD effect, we made corrections to $K^- \pi^+$ 
vs. $K_{S,L}^0 \pi^+ \pi^-$ yields in Section~\ref{section:fit} 
by using results from Ref.~\cite{TQCA}.
The systematic uncertainties due to $r$ is negligible since it is precisely measured.
The systematic uncertainties due to the strong phase $\delta$ are
studied by varying it according to its error.
For $K^- \pi^+ \pi^0$, and $K^- \pi^+ \pi^+ \pi^-$ tag modes, there are no relative strong phase
measurements, so we consider four cases, $\delta = (0^\circ, 90^\circ, 180^\circ, 270^\circ)$,
and take the maximum variations among the four cases as the systematic uncertainty.

The total systematic uncertainties on ($c_i$, $s_i$) and ($c'_i$, $s'_i$) $-$ excluding the systematic uncertainty on $\Delta c_i$ and $\Delta s_i$ that relate the $K^0_S\pi^+\pi^-$ and $K^0_L\pi^+\pi^-$ Dalitz-plot models $-$ are obtained from the quadrature sum of these systematic uncertainties, are shown in Tables~\ref{table:sys_ci} - Table~\ref{table:sys_s'i}.

In the global fit, $\Delta c_i$ and $\Delta s_i$ are constrained using a 
$\chi^2$ term. The errors on $\Delta c_i$ and $\Delta s_i$ are determined by comparing
BaBar, Belle, and CLEO II $D^0 \to K_S^0 \pi^+\pi^-$ Dalitz-plot fit results. 
The constraint on $c$  and $c'$ can be removed with little impact on the result.
The constraint on $s$ and $s'$ can be relaxed to a factor of 4, but cannot be removed entirely, otherwise, the fit does not converge.
To assess our sensitivity to this constraint we consider the following 1) we 
relax the constraint by a factor of 2, {\it i.e.} increase the errors by a factor of 2 and re-fit the data.
2) we fix $\Delta c_i$ and $\Delta s_i$ and re-fit the data (see Table~\ref{table:ci_si}).
The maximum difference for each ($c_i$, $s_i$) and ($c'_i$, $s'_i$) between these fits and the nominal fit is interpreted as the systematic uncertainty. 
An alternate assessment of this systematic uncertainty is the
difference in quadrature of the errors reported for the ``fixed" and ``constrained" fits reported in 
Table~\ref{table:ci_si}. The quadrature average of these two methods is reported as 
the third error on ($c_i$,$s_i$) and ($c'_i$,$s'_i$) in Table~\ref{table:ci_si-final} and Table~\ref{table:c'i_s'i-final}, respectively.

\section{Cross check for model predictions}
\begin{figure}
\centering
\includegraphics[width=8cm]{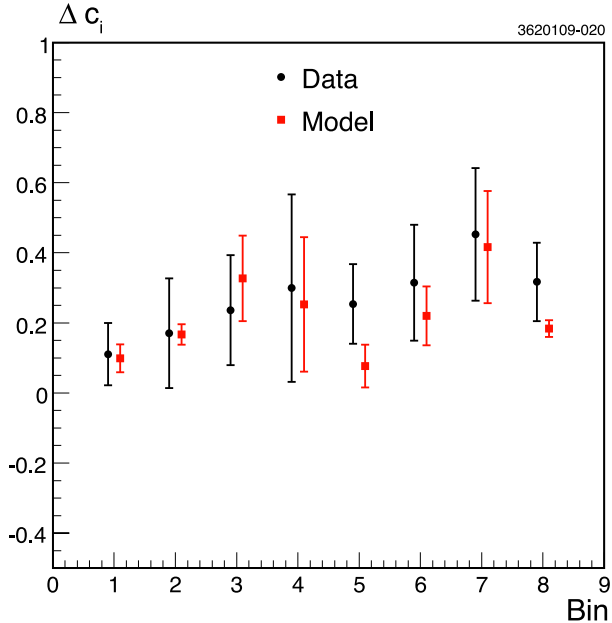}
\caption{Comparison of $\Delta c_i = c_i'-c_i$ between CP-tagged $K_{S/L}^0\pi^+\pi^-$ CLEO-c
data (circles) and predictions from the BaBar model (squares).}
\label{fig:diff_comp}
\end{figure}

Using CP-tagged $K_{S/L}^0\pi^+\pi^-$ samples, we can get $c_i$ and $c_i'$ without any 
correlations, so the differences between $c_i$ and $c_i'$ provide a good test of our
predictions on the differences discussed in Section~\ref{section:diff_KsKl}. The comparison between
measured $\Delta c_i= c_i'-c_i$ and the BaBar model predictions is shown in Fig.~\ref{fig:diff_comp}.
We find good agreement between the data and the results obtained using the BaBar model,
modified to account for the difference between $D^0\to K^0_S\pi^+\pi^-$ and $D^0\to K^0_L\pi^+\pi^-$. 

\section{\boldmath Final Results and Impact on $\gamma/\phi_3$ measurement}
Our final results for $c_i$, $s_i$, $c_i'$ and $s_i'$ are shown in Table~\ref{table:ci_si-final} and 
Table~\ref{table:c'i_s'i-final}, respectively.
The statistical uncertainties dominate for $c_i$ and $s_i$. The systematic uncertainty due to $\Delta c_i$ and $\Delta s_i$ which relate the strong phase difference of $D^0 \to K^0_S\pi^+\pi^-$ and $D^0 \to K^0_L\pi^+\pi^-$ is comparable to $-$ but does not dominate $-$ all other contributions to the total systematic uncertainty.
\begin{table}[t]
\centering
\caption{Fit results for $c_i$ and $s_i$. 
The first error is statistical, the second error is the systematic uncertainty (excluding $\Delta c_i$, $\Delta s_i$), the third error is the systematic uncertainty due to $\Delta c_i$ and $\Delta s_i$ that relate the $K^0_S\pi^+\pi^-$ and $K^0_L\pi^+\pi^-$ Dalitz-plot models .}
\begin{tabular}{ccc}        \hline\hline
 $i$     & $c_i$  & $s_i$  \\\hline
0 & $\hphantom{-}0.743 \pm 0.037 \pm 0.022 \pm 0.013$ & $\hphantom{-}0.014 \pm 0.160 \pm 0.077 \pm 0.045$ \\
1 & $\hphantom{-}0.611 \pm 0.071 \pm 0.037 \pm 0.009$ & $\hphantom{-}0.014 \pm 0.215 \pm 0.055 \pm 0.017$ \\
2 & $\hphantom{-}0.059 \pm 0.063 \pm 0.031 \pm 0.057$ & $\hphantom{-}0.609 \pm 0.190 \pm 0.076 \pm 0.037$ \\
3 & $-0.495 \pm 0.101 \pm 0.052 \pm 0.045$ & $\hphantom{-}0.151 \pm 0.217 \pm 0.069 \pm 0.048$ \\
4 & $-0.911 \pm 0.049 \pm 0.032 \pm 0.021$ & $-0.050 \pm 0.183 \pm 0.045 \pm 0.036$ \\
5 & $-0.736 \pm 0.066 \pm 0.030 \pm 0.018$ & $-0.340 \pm 0.187 \pm 0.052 \pm 0.047$ \\
6 & $\hphantom{-}0.157 \pm 0.074 \pm 0.042 \pm 0.051$ & $-0.827 \pm 0.185 \pm 0.060 \pm 0.036$ \\
7 & $\hphantom{-}0.403 \pm 0.046 \pm 0.021 \pm 0.002$ & $-0.409 \pm 0.158 \pm 0.050 \pm 0.002$ \\
\hline\hline
\label{table:ci_si-final}
\end{tabular}
\end{table}

\begin{table}[t]
\centering
\caption{Fit results for $c'_i$ and $s'_i$. 
The first error is statistical, the second error is the systematic uncertainty (excluding $\Delta c_i$, $\Delta s_i$), the third error is the systematic uncertainty due to $\Delta c_i$ and $\Delta s_i$ .}
\begin{tabular}{ccc}        \hline\hline
 $i$     & $c_i'$  &  $s_i'$ \\\hline
0 & $\hphantom{-}0.840 \pm 0.037 \pm 0.023 \pm 0.014$ & $-0.021 \pm 0.160 \pm 0.080 \pm 0.036$ \\
1 & $\hphantom{-}0.779 \pm 0.071 \pm 0.039 \pm 0.008$ & $-0.069 \pm 0.215 \pm 0.060 \pm 0.047$ \\
2 & $\hphantom{-}0.250 \pm 0.063 \pm 0.029 \pm 0.102$ & $\hphantom{-}0.587 \pm 0.190 \pm 0.072 \pm 0.006$ \\
3 & $-0.349 \pm 0.101 \pm 0.057 \pm 0.092$ & $\hphantom{-}0.275 \pm 0.217 \pm 0.067 \pm 0.058$ \\
4 & $-0.793 \pm 0.049 \pm 0.029 \pm 0.036$ & $-0.016 \pm 0.183 \pm 0.046 \pm 0.042$ \\
5 & $-0.546 \pm 0.066 \pm 0.028 \pm 0.038$ & $-0.388 \pm 0.187 \pm 0.056 \pm 0.072$ \\
6 & $\hphantom{-}0.475 \pm 0.074 \pm 0.026 \pm 0.081$ & $-0.725 \pm 0.185 \pm 0.065 \pm 0.058$ \\
7 & $\hphantom{-}0.591 \pm 0.046 \pm 0.021 \pm 0.011$ & $-0.374 \pm 0.158 \pm 0.059 \pm 0.054$ \\
\hline\hline
\label{table:c'i_s'i-final}
\end{tabular}
\end{table}

\begin{figure}
\centering
\includegraphics[width=5cm]{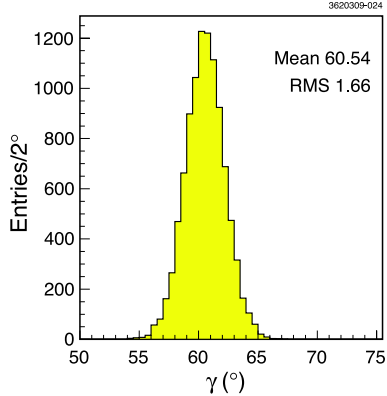}
\caption{Toy Monte Carlo Fit Results for $\gamma/\phi_3$.}
\label{fig:gamma}
\end{figure}

To see the impact of our results on the $\gamma/\phi_3$ measurement, we generate toy Monte Carlo
$B^{\pm}\to \tilde D^0 K^{\pm}$ samples with $\gamma/\phi_3=60^\circ$, $\delta_B=130^\circ$ and $r_B=0.1$. 
The $B^{\pm}\to \tilde D^0 K^{\pm}$ sample 
is large enough so that the statistical uncertainty associated 
with $B$ decays is negligible. We assume the reconstruction efficiency is 100$\%$ and 
that no background is present.
We fit for $\gamma/\phi_3$, $\delta_B$, and $r_B$ 10,000 times 
by sampling $c_i$
and $s_i$ according to their uncertainties and correlations. We find the width of the resulting
$\gamma/\phi_3$ distribution, shown in Fig.~\ref{fig:gamma}, is about $1.7^\circ$. However, a small bias of $0.5^\circ$ is observed,
which is believed to be caused by the unphysical $c_i$ and $s_i$ pairs (617 out of 8000) with $c_i^2 + s_i^2 > 1$. 

Comparing with a model uncertainty of $7^\circ$ for BaBar~\cite{GGSZ_BaBar2} and $9^\circ$
for Belle~\cite{GGSZ_Belle}, great improvement on the $\gamma/\phi_3$ measurement can be achieved 
by using a model-independent approach incorprating CLEO-c's results on the strong phase 
parameters $c_i$ and $s_i$ presented in this article. This will be realized at LHCb where using 
10 fb$^{-1}$ of data a {\it statistical} error on $\gamma/\phi_3$ of $5.5^\circ$ is anticipated \cite{Libby}.
The weight of $B \to \tilde DK, \tilde D \to K^0_S\pi^+\pi^-$ in the
combination of tree-level $\gamma$ measurements at LHCb, which is
predicted to have sensitivity of $1^{\circ}-2^{\circ}$ \cite{akiba}, depends
upon the CLEO-c's results on the strong phase parameters $c_i$ and $s_i$ presented in this article.

Sensitivity to New Physics is obtained through the comparison of $\gamma/\phi_3$ measured directly
in tree-level processes and indirect determinations of $\gamma/\phi_3$.  
One indirect determination, $\gamma/\phi_3=(67^{+5}_{-4})^{\circ}$, arises from the intersection of the $B_{(s)}$ mixing and $\sin 2\beta$ contours in the $(\bar \rho$, $\bar \eta)$ plane \cite{CKMfitter}.
The uncertainty is dominated by the LQCD calculations for mixing \cite{LQCD} 
and are expected to improve.
Another determination of $\gamma/\phi_3$ follows from the unitarity constraint $\gamma = 180^\circ - \alpha - \beta = (70^{+6}_{-5})^{\circ}$. Here the uncertainty is dominated by the determination of $\alpha/\phi_1 = (88^{+6}_{-5})^{\circ}$ from $B\to\pi\pi,\rho\pi,\rho\rho$ \cite{CKMfitter}.


\section{Summary}
In summary, using 818 pb$^{-1}$ of $e^+e^-$ collisions produced at the $\psi(3770)$, we
make a first determination of the strong phase parameters, $c_i$ and $s_i$, in Table~\ref{table:ci_si-final}.
From a toy Monte Carlo study with a large sample of $B^{\pm}\to \tilde D^0 K^{\pm}$ data generated with
$\gamma/\phi_3=60^\circ$, $\delta_B=130^\circ$ and $r_B=0.1$, we find that the decay model uncertainty on
$\gamma/\phi_3$ is reduced to about $1.7^\circ$ due to these new measurements. As a result,
the precision of the $\gamma/\phi_3$ measurement using $B^+\to\tilde D^0 K^+$ decays
will not be limited by the strong interference effects in the $\tilde D^0\to K^0_S\pi^+\pi^-$ decay.
The improved precision in the direct determination of $\gamma/\phi_3$ enabled by this measurement of the strong phase parameters $c_i$ and $s_i$ enhances sensitivity to New Physics through the comparison with indirect determinations of $\gamma/\phi_3$.

\section{Acknowledgments}
We gratefully acknowledge the effort of the CESR staff
in providing us with excellent luminosity and running conditions.
D.~Cronin-Hennessy and A.~Ryd thank the A.P.~Sloan Foundation.
This work was supported by the National Science Foundation,
the U.S. Department of Energy,
the Natural Sciences and Engineering Research Council of Canada, and
the U.K. Science and Technology Facilities Council.

\end{document}